\documentclass[draft]{agujournal2019}
\usepackage{url} 
\usepackage{lineno}
\usepackage{soul}
\usepackage{amsmath}
\usepackage{placeins}
\usepackage{amssymb}
\nolinenumbers
\makeatletter
\def\@oddhead{}
\let\@evenhead\@oddhead
\makeatother
\draftfalse

\journalname{Geophysical Research Letters}

\begin{document}

%
%

\title{Physical Processes Driving Year-to-Year Surface Temperature Variability Changes in a Warming Climate}

%
%




\authors{Rory Basinski-Ferris\affil{1}\ and Ian Eisenman\affil{1}}


\affiliation{1}{Scripps Institution of Oceanography, University of California San Diego}




\correspondingauthor{Rory Basinski-Ferris}{rbasinskiferris@ucsd.edu}


\begin{keypoints}
\item A stochastic moist energy balance model largely reproduces the zonal-mean year-to-year temperature variability change under warming in CESM
\item Weaker anomalous latent heat flux convergence and higher effective heat capacity from sea ice loss reduce mid-to-high latitude variability
\item Tropical and subtropical variability increases due to changes in the local feedback parameter
\end{keypoints}


%
%

%
%


\begin{abstract} 
Under anthropogenic greenhouse forcing, climate models project a decrease in year-to-year temperature variability in the mid-to-high latitudes and an increase in the tropics and subtropics. Here, we examine year-to-year temperature variability in the CESM Large Ensemble in concert with an idealized stochastic moist energy balance model (SMEBM) with effective surface heat capacity and local feedback parameter specified based on CESM. We find that the zonal-mean CESM variability change pattern largely persists after filtering out the impact of El Niño. The SMEBM largely captures this change pattern. In the SMEBM, changes in the mid-to-high latitudes are driven by changes in the effective heat capacity due to sea ice retreat, combined with changes in anomalous moist static energy convergence modulated by the polar-amplified time-mean temperature change. The CESM local feedback parameter evolves to be less stabilizing in the low latitudes, which causes tropical and subtropical variability to increase in the SMEBM.
\end{abstract}

\section*{Plain Language Summary}
Under global warming, the frequency of hot temperature extremes changes because of both warmer time-average temperatures and changes in the amplitude of variability. Climate model simulations demonstrate that in many low-latitude regions, the amplitude of temperature variability increases, whereas outside of this latitude range, it tends to decrease. Here, we present an idealized physical model to explain this pattern. The increase in low latitude variability is linked to a change in how efficiently energy is emitted from the earth to space. The local variability decreases outside this region are attributed to losing sea ice and to the atmosphere redistributing anomalous heat differently.

%
%

%


%
%
%
%

\section{Introduction}

Earth's climate evolves in time due to both external forcing and internal variability. Previous work suggests that greenhouse forcing alters not only the time-mean temperature, but also the amplitude of its internal variability about the mean \cite<e.g.,>{Olonscheck2021Large-scaleCentury,Shi2023ProjectedSimulations,Huntingford2013NoPatterns,Bathiany2018ClimateCountries,Brown2017ChangeWarming,Coquereau2024AnthropogenicSimulations}. Changes in both the mean and the variability modify the probability of a given temperature \cite<e.g.,>{Lewis2017EvolutionTemperatures,vanderWiel2021ContributionExtremes}, such that local increases in variability amplify the increase in hot extremes due to mean warming, while decreases offset it.

The question of how variability changes under greenhouse forcing over a range of timescales has received attention \cite<e.g.,>{Manabe1980OnAtmosphere,Raisanen2002CO2-InducedExperiments,Stouffer2007ChangesTemperature,Huntingford2013NoPatterns,Screen2014ArcticHigh-latitudes,Schneider2015PhysicsVariability,Holmes2016RobustAdvection,Brown2017ChangeWarming,Bathiany2018ClimateCountries}. In particular, on year-to-year timescales, simulations using models from the Coupled Model Intercomparison Project Phase 5 (CMIP5) and Phase 6 (CMIP6) under high-emissions scenarios (RCP8.5 or SSP5-8.5) largely project increased low-latitude and decreased mid-to-high-latitude temperature variability \cite{Olonscheck2021Large-scaleCentury,Shi2023ProjectedSimulations,Huntingford2013NoPatterns,Coquereau2024AnthropogenicSimulations}. The high-latitude decrease, especially in the Northern Hemisphere, and the low-latitude increase over land are particularly robust across models \cite<e.g.,>{Olonscheck2021Large-scaleCentury}.

The relative role of different proposed mechanisms governing these year-to-year internal variability changes is debated. Many studies have attributed mid-to-high-latitude variability decreases to sea ice loss \cite<e.g.,>{Stouffer2007ChangesTemperature,Huntingford2013NoPatterns,Brown2017ChangeWarming,Olonscheck2021Large-scaleCentury,Coquereau2024AnthropogenicSimulations}, which allows the atmosphere to interact with the large heat capacity of the ocean. Another suggested contributor is the flattening of the meridional temperature gradient under polar-amplified warming, which has been found to reduce daily-to-monthly variability via thermal advection \cite{Screen2014ArcticHigh-latitudes,Schneider2015PhysicsVariability,Holmes2016RobustAdvection}. On year-to-year timescales, on the other hand, \citeA{Coquereau2024AnthropogenicSimulations} argue that temperature gradient flattening is of secondary importance, whereas \citeA{Olonscheck2021Large-scaleCentury} and \citeA{Shi2023ProjectedSimulations} suggest that this mechanism may be an important contributor to year-to-year variability decreases. Low-latitude year-to-year variability increases have previously been attributed to local drying of land surfaces \cite<e.g.,>{Olonscheck2021Large-scaleCentury,Shi2023ProjectedSimulations} and increases in longwave radiation from the mean cloud cover \cite{Shi2023ProjectedSimulations}, or alternatively to changes in the El Ni\~no--Southern Oscillation (ENSO) \cite{Coquereau2024AnthropogenicSimulations}.

Previous mechanistic work on year-to-year surface temperature variability changes has largely used comprehensive climate models, in which the many active processes are difficult to disentangle. An idealized model offers a complementary approach by sacrificing the realism of comprehensive models in exchange for clearer interpretability, such that different ingredients can be switched on or off to isolate their contributions. The use of simpler models as a link between comprehensive model simulations and physical understanding is long established in climate dynamics \cite<e.g.,>{Held2005TheModeling,Vallis2016GeophysicalWhy,Jeevanjee2017AHierarchies}. 

A prominent class of idealized climate models is Budyko-Sellers energy balance models (EBMs) \cite{Budyko1969TheEarth,Sellers1969ASystem}, which simulate the zonal-mean temperature driven by top-of-the-atmosphere fluxes and meridional heat transport. This transport is typically parameterized as Fickian diffusion of either near-surface temperature \cite<``dry" EBMs; e.g., >{North1975AnalyticalTransport,Held1974SimpleIcecaps} or moist static energy \cite<moist EBMs (MEBMs); e.g.,>{Flannery1984EnergyEnergy}. MEBMs represent both sensible and latent heat transport and have successfully recreated features of the time-mean forced climate response simulated by comprehensive climate models \cite<e.g.,>{Hwang2010IncreasingWarming, Rose2014TheUptake,Roe2015ThePredictability,Siler2018InsightsModel,Bonan2018SourcesChange,Merlis2018SimpleModels,Armour2019MeridionalDiffusion,Beer2022RevisitingChange}. The MEBMs adopted in these previous studies do not have a representation of internal climate variability. On the other hand, some previous studies have used dry EBMs with imposed weather noise to represent internal variability in the present climate \cite{Kim1991SurfaceModel,Kim1992SeasonalSystem,Kim1996ComparisonsModels,Rypdal2015SpatiotemporalModels}, following Hasselmann's stochastic climate perspective \cite{Hasselmann1976StochasticTheory}. 

Here, we add imposed weather noise to a MEBM to investigate future changes in internal variability. We examine year-to-year zonal-mean temperature variability in the Community Earth System Model (CESM) Version 1 Large Ensemble, in concert with this stochastic MEBM (SMEBM). We inform the SMEBM with relevant parameters from CESM under anthropogenic forcing. We demonstrate that the SMEBM can largely capture the variability changes in CESM, and we isolate the ingredients needed to reproduce these variability changes in each region.

\section{Methods}
\subsection{CESM-LE Data}
We analyze data from the 40-member CESM Version 1 Large Ensemble \cite{Kay2015TheVariability}, where the forcing is historical from 1920 until 2005 and then follows the Representative Concentration Pathway (RCP) 8.5 until 2100. We compute the variation across ensemble members in each year. We use this as a continuous measure of year-to-year internal variability of any individual ensemble member \cite<following e.g.,>{Olonscheck2021Large-scaleCentury}, which is consistent with the quasi-ergodic assumption \cite{Olonscheck2017ConsistentlySimulations}. Hence, variability is calculated as
\begin{align}
    \sigma_{ens}(t)=\sqrt{\frac{1}{n-1}\sum_{i=1}^n\left(T(i,t)-\left<T\right>_{ens}(t)\right)^2},  \textrm{  with} \left<T\right>_{ens}(t)=\frac{1}{n}\sum_{i=1}^n T(i,t),
    \label{eq:ens_variance} 
\end{align}
where $T(i,t)$ is the annual-mean near-surface temperature of the $i^{\textrm{th}}$ ensemble member at time $t$ and $n=40$ is the total number of ensemble members. This yields a standard deviation that can be evaluated at each year.

\subsection{ENSO filtering method}
ENSO variability is not represented in an energy balance framework, but it may drive surface temperature variability changes. Thus, we remove its influence from the CESM results using a regression method \cite<similar to e.g.,>{Liu2026ENSOWarming,Robock1995TheObservations,Angell2000Tropospheric19581998,Chiang2004AnalogousVariability}. 

First, we take the target variable (e.g., surface air temperature) at monthly-mean time resolution in a matrix $A$ and compute anomalies $A'$ by subtracting the ensemble mean from the total field. Next, we compute the Oceanic Niño Index (ONI) as the area-weighted monthly sea surface temperature anomaly (from the ensemble mean) spatially averaged over the Niño 3.4 region. We filter the index with a 4th-order Butterworth bandpass filter between 2 and 7 years to remove variations outside of the typical ENSO periodicity (denoted $m$ at 0-lag). Finally, we regress the zonal-mean of $A'$ against $m$ separately at each latitude to find the linear slope vector $s$. The ENSO-filtered field is defined as $A'_{\textrm{ENSOfiltered}}=A'-sm$. 

As noted by \citeA{Liu2026ENSOWarming}, this approach assumes that the relationship between the variable ($A$) and ENSO variability is both instantaneous and linear. To account for potential lagged or non-linear relationships, we also use an alternative version, replacing the linear regression with a machine learning approach. We train a Multilayer Perceptron (MLP) at each latitude to predict $A'$ using the bandpass-filtered ONI with lags of 0, 2, 4, 6, 8, 10, and 12 months as input features (denoted $m_{\textrm{lags}}$). The MLP is implemented via scikit-learn \cite{Pedregosa2011Scikit-learn:Python} with two hidden layers (of size 12 and 8), a ReLu activation function, and an $L^2$ regularization penalty to prevent overfitting ($\alpha=0.2$). The filtered residuals are calculated as: $A'_{\textrm{ENSOfiltered, MLP}}=A'-\textrm{MLP}(m_{\textrm{lags}})$, where $\textrm{MLP}(\cdot)$ represents the trained non-linear mapping.

\subsection{Stochastic moist energy balance model} \label{subsec:SEBM}
We begin with a standard deterministic moist energy balance model \cite<MEBM; e.g.,>{Rose2014TheUptake,Roe2015ThePredictability,Siler2018InsightsModel,Bonan2018SourcesChange,Armour2019MeridionalDiffusion,Beer2022RevisitingChange}, defined in terms of perturbations $T'(x,t)$ around a mean temperature climatology ($T_0(x)$) and neglecting changes in ocean heat flux convergence and deep ocean heat uptake:
\begin{align}
C(x) T'(x,t)_t &= \underbrace{D \nabla^2 h'(x,t)}_{\text{(a)}} + \underbrace{\lambda(x)T'(x,t)}_{\text{(b)}} +\underbrace{F}_{\text{(c)}}, \textrm{ with} \label{eq:mEBM} \\
h(T)&=c_p T+L\underbrace{\frac{\epsilon r e_0}{p} \exp\left(\frac{aT}{b+T}\right)}_{q}\textrm{, and } \label{eq:mEBM2}\\
h'(x,t)&=\frac{dh(T)}{dT}\Bigr|_{T_0(x)} T'(x,t).
\label{eq:MSE}
\end{align}

In Equation \eqref{eq:mEBM}, term (a) is the convergence of the meridional transport (where the Laplacian accounts for spherical geometry in latitude coordinates), term (b) is the anomalous top-of-the-atmosphere (TOA) emitted radiation, and term (c) is an optional radiative forcing. All parameters are as in \citeA{Beer2022RevisitingChange}. Moist static energy (MSE), denoted as $h$, is defined in Equation \eqref{eq:mEBM2}, with $q$ denoting the specific humidity. Anomalies in MSE are equal to the first-order Taylor series approximation as specified in Equation \eqref{eq:MSE} \cite<following, e.g.,>{Merlis2018SimpleModels}. See the derivation of the model in Supplementary Text S1. 

In this study, we develop a stochastic MEBM (SMEBM), which we define as
\begin{equation}
T'(x,t)_t = 
\frac{1}{C(x)}\left[\underbrace{D \nabla^2 h'(x,t)+\lambda(x)T'(x,t)}_{\textrm{RHS of deterministic equation}} + \underbrace{\sigma \xi(x,t)}_{\textrm{noise term}}\right]
\label{eq:Stochastic-EBM} 
\end{equation}
where $\xi(x,t)$ is space-time white noise such that $\mathbb{E}[\xi(x,t)\xi(x',t')]=\delta(x-x')\delta(t-t')$. For simplicity, the intensity of the noise at each point, $\sigma$, is taken to be spatially uniform \cite<following, e.g.,>{Kim1992SeasonalSystem}.
As discussed in Text S1, the noise intensity $\sigma$ is tuned to match the early-period variability and then left unchanged in all simulations. This noise is inserted into the relationship between variations in surface temperature and TOA energy flux, which can be thought of as arising from various weather-driven processes in the climate system such as cloud fluctuations and non-local effects on the TOA radiation. 

Temperature variability in the SMEBM is calculated by running simulations of a 100-member ensemble for 150 years. We then coarsen to annual averages and take the ensemble standard deviation at every year following Equation (1), and we average the results over the final 50 years of the 150-year simulations.

\FloatBarrier
\begin{figure*}[h!]
\noindent\includegraphics[width=\textwidth,angle=0]{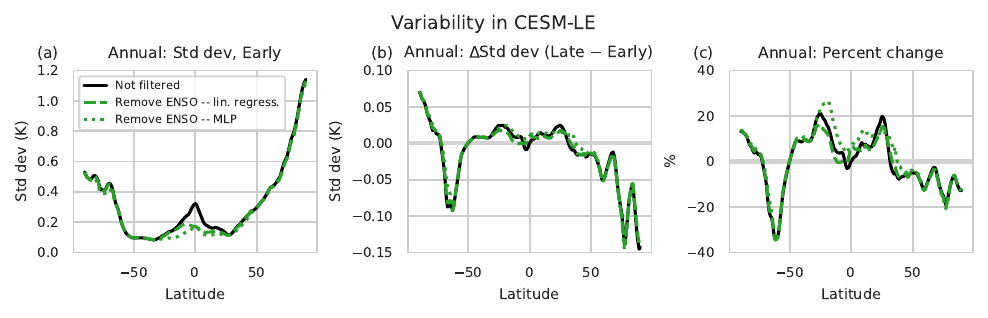}\\
  \caption{Ensemble standard deviation in CESM (Equation \eqref{eq:ens_variance}) averaged over either the early (2000-2025) or late (2075-2100) 21st century. (a): Variability in the early period. (b): Variability change between the late and early periods. (c): As in panel (b), but expressed as a percentage of the early period value. In all panels, the raw variability is shown (black) as well as the variability after filtering out ENSO via linear regression (green, dashed) and via the MLP (green, dotted).} \label{fig:CESMLE}
\end{figure*}

\subsection{Parameter estimation from CESM} \label{sec:parameter_estimation}
We compute the effective surface heat capacity and local feedback parameter from the CESM runs to inform the SMEBM.

Following \citeA{Kim1991SurfaceModel,Kim1992SeasonalSystem}, the effective surface heat capacity ($C(\textrm{lat},\textrm{lon},t)$) is computed by assuming that the ocean, land, and sea ice each have a set heat capacity, and weighting between components based on the land mask and the ensemble-mean ice fraction. We set the mixed layer heat capacity to 8.74 W yr m$^{-2}$ K$^{-1}$, and set the sea ice and land capacities as $1/65$ and $1/600$ relative to the ocean, respectively \cite<the same values as adopted in>{Kim1992SeasonalSystem}. We use the heat capacity of sea ice for the Greenland and Antarctic ice sheets. See Texts S1 and S2 for further discussion.

We compute the local feedback parameter from internal variability by regressing zonal-mean TOA flux anomalies against zonal-mean surface temperature anomalies, after applying the ENSO filtering method and coarsening to an annual mean timescale.
The slope of this regression at every location is the feedback parameter that damps non-ENSO anomalies. The regression is performed separately to diagnose the feedback parameter near the beginning of the 21st century (using data from years 2000-2025) or near the end of the 21st century (using data from years 2075-2100). For analysis only, we compare the variability feedback parameter with the forced feedback parameter decomposed with a kernel analysis using the ClimKern package \cite{Janoski2025ClimKernFeedbacks} (Figure S8).

\section{Variability in the CESM Large Ensemble}
The zonal-mean variability in the early period (defined as the mean over years 2000-2025) in CESM peaks at the poles with more variability in the Northern Hemisphere than the Southern Hemisphere. There is a secondary peak in the tropics centered at the equator (Figure 1a). Applying both the linear and non-linear ENSO filtering methods to surface air temperature demonstrates that the tropical peak is primarily ENSO-driven.

The change in the temperature standard deviation at the end of the 21st century (defined as the mean over years 2075-2100) compared to the early period approximately follows the trends in CMIP5 and CMIP6 models discussed in previous studies \cite{Bathiany2018ClimateCountries,Olonscheck2021Large-scaleCentury,Coquereau2024AnthropogenicSimulations,Shi2023ProjectedSimulations,Huntingford2013NoPatterns,Brown2017ChangeWarming}: the zonal-mean variability increases in the subtropics and tropical low latitudes and decreases in the mid-latitudes. Polar variability decreases in the Northern Hemisphere and increases in the Southern Hemisphere, although there are local decreases over the Southern Ocean (Figure 1b). While we focus on the zonal-mean change here, we note that the strongest low-latitude variability increases are over land (Figure S1). The change in variability is relatively insensitive to ENSO filtering aside from small impacts in the tropics and subtropics, indicating that these changes are not primarily ENSO-driven. These changes are broadly consistent with other time period choices (see Figure S2 for the year-to-year zonal-mean variability change over time).

While the largest absolute standard deviation changes occur in the mid-to-high latitudes of both hemispheres, the low latitudes have a substantial relative increase ($\sim 20\%$), despite a small absolute change ($\sim 0.05$K). This relative change is of the same magnitude as the change in the northern high latitudes (Figure 1c).

\FloatBarrier

\begin{figure*}[h!]
  \noindent\includegraphics[width=\textwidth,angle=0]{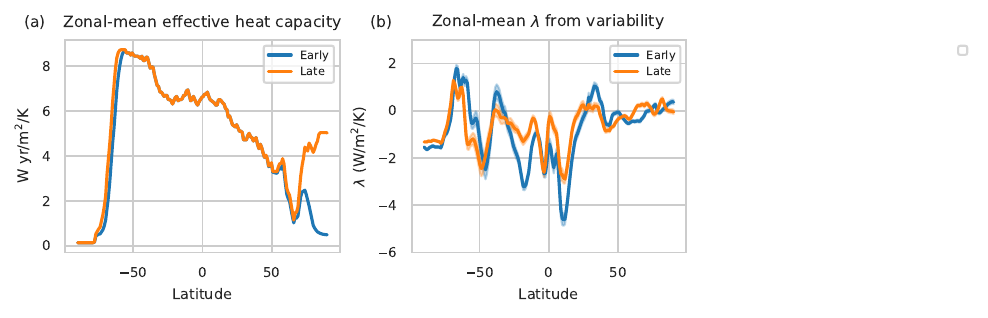}\\
  \caption{Parameters fitted from CESM. (a): The zonal-mean effective heat capacity using assigned values for the ocean, land, and sea ice. (b): The zonal-mean feedback parameter found via regression (Section \ref{sec:parameter_estimation}) and used in the SMEBM.} \label{fig:parameters}
\end{figure*}

\section{Variability in the stochastic energy balance model}
Here, we discuss the early-period variability by simulating Equation \ref{eq:Stochastic-EBM} using early 21st-century parameters derived from CESM, and computing variability using Equation \ref{eq:ens_variance}. 
To isolate drivers of future variability change, we update these parameters to their late 21st-century values and compare the resulting variability against the early period.

\subsection{The effect of ensemble-mean temperature}

Temperature variability depends on the latitudinally-varying ensemble-mean temperature ($T_0(x)$) because the transport convergence, $D \nabla^2 h'(x,t)$, depends on $T_0(x)$ according to Equation \eqref{eq:MSE}. This dependence on $T_0(x)$ is entirely through the latent energy flux convergence, rather than the sensible energy flux convergence.
Using the product rule on the Laplacian of Equation \eqref{eq:MSE}, we can write:
\begin{align}
    D\nabla^2h'(x,t)&=D\left[\frac{dh}{dT}\Bigr|_{T_0(x)}\nabla^2T'(x,t)+2\left(\nabla\frac{dh}{dT}\Bigr|_{T_0(x)}\right)\nabla T'(x,t)+T'\nabla^2\frac{dh}{dT}\Bigr|_{T_0(x)}\right] \\
    &=\underbrace{D_{\textrm{eff}}(x)\nabla^2T'(x,t)}_{\text{(i)}}+\underbrace{2D\left(\nabla\frac{dh}{dT}\Bigr|_{T_0(x)}\right)\nabla T'(x,t)}_{\text{(ii)}}+\underbrace{\lambda_{\textrm{eff}}(x)T'(x,t)}_{\text{(iii)}}.
    \label{eq:transportvariancerewrite}
\end{align}
where $D_{\textrm{eff}}(x)=D\frac{dh}{dT}\Bigr|_{T_0(x)}$ and $\lambda_{\textrm{eff}}(x)=D\nabla^2\frac{dh}{dT}\Bigr|_{T_0(x)}$.

\begin{figure*}[h!]
  \noindent\includegraphics[width=\textwidth,angle=0]{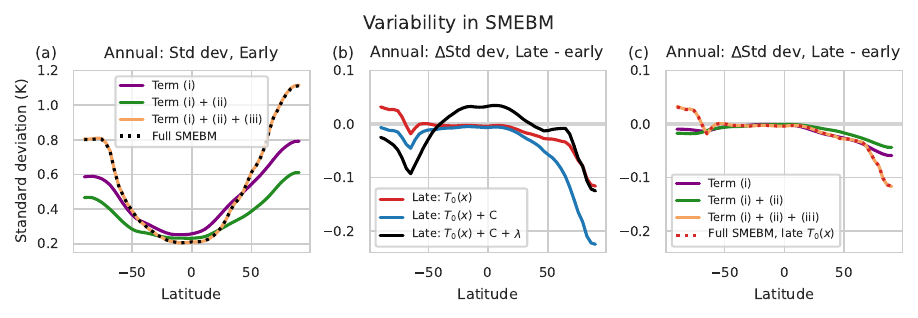}\\
  \caption{Variability in the SMEBM in (a) the early period, (b) the difference between the late and early periods, and (c) the difference between the late and early periods driven only by transport convergence. Decompositions into the transport convergence terms in Equation \ref{eq:transportvariancerewrite} are shown on panels (a) and (c). Decomposition into the effect of changing background temperature, heat capacity, and the feedback parameter are shown on panel (b).} \label{fig:sebm}
\end{figure*}

In the early period, the transport convergence of temperature anomalies via Equation \ref{eq:transportvariancerewrite} results in polar amplification of the temperature standard deviation (Figure \ref{fig:sebm}a). This spatial structure emerges largely through the combined effects of terms (i) and (iii). Term (i) states that the effective diffusivity ($D_{\textrm{eff}}(x)$) acting on a temperature anomaly is larger where the sensitivity of MSE to temperature is higher. Because of the exponential form of the Clausius-Clapeyron relation, $D_{\textrm{eff}}(x)$ is substantially larger in the tropics than in the high latitudes. Thus, anomalies are smoothed more through term (i) at lower latitudes. Term (iii) acts as an effective damping term such that temperature anomalies are damped if the local sensitivity of $h$ to $T$ exceeds that of neighboring regions ($\lambda_{\textrm{eff}}(x)<0$), and amplified if the sensitivity is lower ($\lambda_{\textrm{eff}}(x)>0$). This means that transport converges temperature perturbations into local areas where they cause smaller MSE perturbations. Consequently, polar variability is amplified via this term because the sensitivity there is lower than the neighboring lower latitudes. Taken together, Equation \ref{eq:transportvariancerewrite} results in polar-peaked variability during the early period due to the preferential convergence of anomalies at the poles. Note that polar-amplified variability was similarly described in a 3-box model in \citeA{ClevelandStout2023FingerprintingModels}, where it was partially attributed to a similar mechanism of preferential transport of energy from the tropics to the extratropics due to the MSE gradient, along with the geographic isolation of the extratropics. Note also that other factors in the SMEBM (e.g., spatially-varying heat capacity and feedback parameter) contribute to the polar-peaked variability in Figure \ref{fig:sebm}a, but the transport mechanism alone is sufficient to produce some polar amplification.

Here, we are primarily concerned with the \textit{change} in temperature variability driven by a changed ensemble-mean temperature $T_0(x)$ due to anthropogenic forcing (Figure S7). To isolate this effect, we compare control simulations to those where $T_0(x)$ is updated to its end of the 21st century value in each of terms (i), (ii), or (iii) in Equation \ref{eq:transportvariancerewrite}. The net effect of changed transport convergence is plotted as the red line in Figure \ref{fig:sebm}b, while the decomposition into terms (i)-(iii) is shown in Figure \ref{fig:sebm}c.

Warming increases $D_{\textrm{eff}}(x)$ (term (i)), leading to an overall decrease in variability. Due to the non-linearity of the Clausius-Clapeyron relation, $D_{\textrm{eff}}(x)$ increases more at low latitudes than at high latitudes. The CESM polar-amplified time-mean warming reduces this latitudinal difference in $D_{\textrm{eff}}(x)$ change but does not overcome it. Nevertheless, with the late 21st century polar-amplified warming pattern, term (i) drives a greater decrease in variability in the high latitudes than the low latitudes (Figure \ref{fig:sebm}c, purple line). This is largely because temperature variability is more sensitive to changes in $D_{\textrm{eff}}(x)$ where the early-period variability was high. This combines with the larger increase in $D_{\textrm{eff}}(x)$ at the high latitudes than would occur with uniform warming and conspires to lower high latitude variability, especially in the Northern Hemisphere.

Term (iii) further reduces high-latitude variability under polar-amplified warming, especially in the Northern Hemisphere (compare the yellow and green lines in Figure \ref{fig:sebm}c). This is due to much larger time-mean warming north of 75$^\circ$N than elsewhere, resulting in a less positive $\lambda_{\textrm{eff}}(x)$ than in the early period. In the early period, polar temperature variability was amplified via term (iii) because MSE sensitivity was much lower than in neighboring regions. Under warming, this amplification weakens because the sensitivity is more similar to neighboring regions due to the larger change in $T_0(x)$ near the pole than elsewhere. In the Southern Hemisphere, variability decreases via term (iii) at $\sim 70^\circ$S, aligning with a local peak in mean warming that is expected to be associated with sea ice loss around the coast of Antarctica (not shown). Crucially, these results depend on the simulated time-mean warming pattern, as well as the early-period temperature pattern, which together determine the change in $\lambda_{\textrm{eff}}(x)$ at each location. If there were instead spatially-uniform or only weakly polar-amplified time-mean warming added to the CESM early-period temperature, then high latitude variability would increase via term (iii), as the non-linearity of the MSE function due to the Clausius-Clapeyron relation would increase $\frac{dh}{dT}$ more at lower latitudes than at higher latitudes (Figure S4).

Polar-amplified temperature variability via transport convergence can be thought of as somewhat analogous to the more familiar polar-amplified time-mean warming, which is driven in part by MSE transport \cite<e.g.,>{Alexeev2005PolarFeedbacks,Held2006RobustWarming,Merlis2018SimpleModels}.
In both cases, a perturbation (stochastic or deterministic) to the background profile $T_0(x)$ is transported down the mean MSE gradient, causing amplified anomalies at the poles. Our results show reduced polar amplification of variability with a shallower equator-to-pole gradient (here, driven by polar-amplified time-mean warming in CESM). This mirrors the smaller deterministic polar amplification found in climates with shallower background gradients (see vertical profiles in 
\citeA[their Figure 4b]{Merlis2018SimpleModels} or horizontal profiles in \citeA[their Figure 1c]{Williams2026State-DependenceGCM}). Similarly, a spatially-uniform warming of $T_0(x)$ leads to greater polar amplification of variability in the SMEBM (Figure S4). This is comparable to enhanced deterministic polar amplification in climates with uniformly warmer time-mean temperature profiles (see horizontal profiles in 
\citeA[their Figure 4b]{Merlis2018SimpleModels} or vertical profiles in \citeA[their Figure 1c]{Williams2026State-DependenceGCM}).

\FloatBarrier

\subsection{The effect of heat capacity and feedback parameter changes}
In the absence of diffusive transport, each latitude (e.g., $x_0$) is an Ornstein-Uhlenbeck process of the form 
\begin{equation}
    C(x_0)dT(x_0,t)=\lambda(x_0) T(x_0,t) dt+\sigma dW(t),
\end{equation}
such that the variance of near-surface temperature scales as
\begin{align}
    \textrm{Var}(T(x_0,t))=-\frac{\sigma^2}{2\lambda(x_0) C(x_0)}\left(1-e^{2\lambda(x_0) t/C(x_0)}\right),
    \label{eq:var_ou}
\end{align}
with time-constant parameters $\lambda(x_0)$, $\sigma$, and $C(x_0)$. Thus, temperature variability scales inversely with both $\lambda(x_0)$ and $C(x_0)$, meaning a larger heat capacity $C(x_0)$ or a more stabilizing local feedback $\lambda(x_0)$ results in decreased variability. The presence of $C(x_0)$ and $\lambda(x_0)$ in the denominator of Equation \eqref{eq:var_ou} also indicates that the change in variability is more sensitive to a given parameter shift if the baseline of that parameter was smaller. We test our expectations regarding the effects of the surface effective heat capacity and feedback parameter in the SMEBM (which includes diffusive transport) by changing these parameters, in addition to the changed background temperature $T_0(x)$ discussed above.

First, we consider the surface effective heat capacity, $C(x)$. In the early period, this parameter is larger in the lower latitudes than the polar latitudes (Figure \ref{fig:parameters}a). This is due to the presence of year-round open ocean at low latitudes, contrasted with sea ice (with a much lower heat capacity) covering the Arctic and Southern Oceans in winter. This latitudinal structure of $C(x)$ contributes to the polar-peaked variance in the climatology, because lower surface effective heat capacity at the poles increases the local variability, as shown in Equation \ref{eq:var_ou}. 

Compared to the early period, the late 21st-century heat capacity is larger at the high latitudes due to the loss of sea ice (Figures \ref{fig:parameters}a, S3, and S9). We examine the effect of changed heat capacity due to anthropogenic forcing on variability by simulating Equation \eqref{eq:Stochastic-EBM} with late 21st-century $C(x)$ values (in addition to changed $T_0(x)$ in the transport convergence term) in comparison to the early-period variability (the blue line in Figure \ref{fig:sebm}b). We find that the increased heat capacity due to sea ice loss lowers high latitude variability. This is especially true in the Northern Hemisphere where the variability decrease due to changing $C(x)$ is largest in the domain and of similar magnitude to the decrease driven by changing $T_0(x)$. In the Southern Hemisphere, the variability decrease peaks around $70^\circ$S due to local sea ice loss, which aligns with the variability decrease driven by the $T_0(x)$ warming peak. 

Next, we consider the feedback parameter, $\lambda(x)$. The feedback parameter diagnosed from CESM has more stabilizing regions located in the low latitudes. This also contributes to the polar-peaked variability in the early period (not shown). Under anthropogenic forcing, the feedback parameter becomes less stabilizing, especially in the low latitudes (Figure \ref{fig:parameters}b). For interpretation, we compare the feedback parameter regressed from variability and used in the SMEBM (Figure \ref{fig:parameters}b) with the feedback parameter as it is alternatively found from the forced ensemble-mean response (Figure S8). The two feedback parameters are qualitatively similar, although quantitatively different. For example, the forced feedback similarly becomes less stabilizing in the low latitudes by the end of the 21st century. Our kernel analysis shows that this change in the low latitudes is primarily driven by the cloud feedback (Figure S8, compare the blue and dashed orange lines). This shift is more linked to the shortwave than the longwave cloud component  (Figure S8, compare the dashed and dotted orange lines).

We examine the effect of changed $\lambda(x)$ due to anthropogenic forcing in the SMEBM by simulating Equation \eqref{eq:Stochastic-EBM} using the late 21st-century local feedback parameter, in addition to the late 21st-century $T_0(x)$ and $C(x)$. Comparing this to the early-period variability, we find that variability increases in the low latitudes (Figure \ref{fig:sebm}b, black line). This is driven by the less stabilizing feedback parameter, as expected from Equation \ref{eq:var_ou}. In the SMEBM, this is the only factor that causes an increase in low latitude variability. We note that the variability also increases in the mid-to-high northern latitudes, relative to the simulations with only updated $T_0(x)$ and $C(x)$ (Figure \ref{fig:sebm}b, compare the black and blue lines), which is expected to be due to transport from the low latitudes. However, there is still a net decrease in variability in this region. The southern mid-to-high latitudes exhibit decreased variability due to a more stabilizing variability feedback parameter over the Southern Ocean, although the robustness of this is uncertain given that the forced feedback parameter shows limited change over this region.

\section{Discussion}
Anthropogenic forcing is expected to change not only the time-mean climate, but also the internal climate variability about it. Under warming, large ensembles of comprehensive climate models project increases in year-to-year surface temperature variability in the low latitudes and decreases in the mid-to-high latitudes. In the present study, we examined the changes in zonal-mean year-to-year surface temperature variability in CESM, in concert with the development and analysis of an idealized stochastic moist energy balance model (SMEBM) to investigate physical mechanisms.

We find that the mid-to-high-latitude variability decrease in the SMEBM arises from changes in the time-mean climate, which modulates latent heat transport convergence, along with sea ice loss, which increases the effective surface heat capacity. The latter is consistent with previous studies \cite<e.g.,>{Stouffer2007ChangesTemperature,Huntingford2013NoPatterns,Brown2017ChangeWarming,Olonscheck2021Large-scaleCentury,Coquereau2024AnthropogenicSimulations}, although our results suggest sea ice loss is not the dominant driver of variability changes (Figure 3b) \cite<in contrast with>{Coquereau2024AnthropogenicSimulations}. 
The first mechanism implies that changes in the time-mean temperature alone cause a substantial decrease in variability. This mechanism differs notably with previous discussions of a weakened time-mean meridional temperature gradient causing reduced year-to-year temperature variability because advected anomalous air masses carry less extreme temperatures \cite{Olonscheck2021Large-scaleCentury,Shi2023ProjectedSimulations}. In the SMEBM, the mean temperature instead affects variability through the latent heat transport, which depends nonlinearly on temperature. In particular, a warmer background climate increases the effective diffusivity ($D_{\textrm{eff}}(x)$, Equation \eqref{eq:transportvariancerewrite}), which more strongly damps temperature anomalies. Additionally, the spatial pattern of time-mean warming, including polar amplification in the Northern Hemisphere, alters $\lambda_{\textrm{eff}}(x)$ (Equation \eqref{eq:transportvariancerewrite}), reducing anomalous energy convergence.

Under the warming considered here, both mechanisms -- the latent heat flux convergence mechanism proposed in the present study and the thermal advection mechanism proposed previously -- imply reduced variability in the mid-to-high-latitude Northern Hemisphere. However, in climates with weak polar amplification, the two mechanisms predict changes of opposite sign: the SMEBM predicts an increase in variability (Figure S4), whereas the meridional temperature gradient decreases. This suggests that the SMEBM transport mechanism is an additional pathway by which sufficiently polar-amplified time-mean warming reduces variability in the mid-to-high latitudes. Note that the present study does not rule out the thermal advection mechanism on year-to-year timescales, because the constant noise intensity assumption excludes it by construction from the SMEBM (Text S1).

In the low latitudes, we find that changes in ENSO are not the primary driver of the year-to-year variability increases, in contrast to \citeA{Coquereau2024AnthropogenicSimulations}, because the increases persist after the ENSO signal is statistically removed using both an established linear method \cite<e.g.,>{Liu2026ENSOWarming} and a nonlinear, lagged machine learning method. In the SMEBM, the low-latitude variability increases instead arise because the local feedback parameter becomes less stabilizing, so that temperature anomalies are damped less effectively. Because the feedback parameter was estimated from CESM after statistically removing ENSO from both surface temperature and TOA fluxes, its change is not expected to reflect an indirect influence of ENSO.

To the extent that the change in the variability feedback is consistent with the change in the forced feedback (Figure S8), this less stabilizing low-latitude feedback agrees with previous work on feedback changes in transiently forced simulations. These studies attribute the evolving feedback to the changing pattern of sea surface warming (or ocean heat uptake), which reduces lower-tropospheric stability, decreases low-cloud cover, and makes the shortwave cloud feedback increasingly positive \cite<e.g.,>{Ceppi2017RelationshipBudget,Rose2016TheSensitivity}, relevant to the broader pattern effect discussion \cite<e.g.,>{Bloch-Johnson2024TheProtocol}. Our results suggest that these feedback changes, previously discussed in the context of the time-mean forced response, may also affect year-to-year variability.

A number of caveats are associated with this work. First, the investigation was designed to explain a single model (CESM), so the identified processes may not carry over to other CMIP models, some of which have been the focus of previous studies. Second, because the SMEBM is highly idealized and its results are sensitive to uncertain fitted parameters, our results should be interpreted qualitatively, as an illustration of physical mechanisms, rather than quantitatively. The feedback parameter is a particular source of uncertainty. Assuming a purely local structure in the regression, as done here, can bias feedback estimates relative to approaches that also account for nonlocal feedbacks \cite<e.g.,>{Bloch-Johnson2020SpatialRegression}. Feedback estimates derived from variability are also sensitive to choices of lag and timescale (e.g., monthly versus annual averaging) \cite<e.g.,>{Lutsko2018WhatSensitivity,Proistosescu2018RadiativeImbalance}. \citeA{Proistosescu2018RadiativeImbalance} showed that zero-lag regression, as used here, can be affected by regression dilution when temperature variability is driven directly by stochastic TOA fluxes. This suggests that our variability-based feedback estimates may be biased. However, the less stabilizing low-latitude feedback by the end of the 21st century is also evident in the forced feedback parameter, which is not subject to the same lag and regression-dilution issues.

Despite these limitations, the SMEBM makes the mechanisms driving year-to-year variability changes in CESM interpretable by simplifying much of the climate system. This study adopts an idealized framework for understanding changes in internal variability under anthropogenic forcing, analogous to the use of energy balance models to understand the time-mean climate response. The results from the SMEBM suggest a novel mechanism by which sufficiently polar-amplified time-mean warming reduces mid-to-high-latitude variability through the temperature dependence of latent heat transport, while also highlighting that less stabilizing local feedbacks may drive low-latitude variability increases.

\section*{Open Research Section}
The output of the CESM Large Ensemble is openly available, including on NSF NCAR Geoscience Data Exchange (doi: 10.5065/D6J101D1). The code used to generate the results in the paper is available at \url{https://github.com/rbasinskiferris/SMEBM} and will be archived on Zenodo by the time of publication.

\section*{Conflict of Interest}
The authors declare there are no conflicts of interest for this manuscript.

\acknowledgments
This project is supported by Schmidt Sciences. We acknowledge high-performance computing support from the Derecho system (doi: 10.5065/qx9a-pg09) provided by the NSF National Center for Atmospheric Research (NCAR), sponsored by the National Science Foundation. Without implying their endorsement, we thank Fabrizio Falasca, Maria Rugenstein, Nick Lutsko, and Kyle Armour for helpful discussions. We also thank Kyle Armour for providing MEBM code. AI tools (versions of Claude and Gemini) were used when developing some of the analysis code used in this work and to assist stylistic edits of the manuscript draft to improve succinctness. The authors are fully responsible for the scientific content and accuracy of the paper.

%
%
\bibliography{references}

%
%
%
%
%

\end{document}


%
%


\title{Supporting Information for ``Physical Processes Driving Year-to-Year Surface Temperature Variability Changes in a Warming Climate"}
%
%

%
%

\authors{Rory Basinski-Ferris\affil{1}\ and Ian Eisenman\affil{1}}


\affiliation{1}{Scripps Institution of Oceanography, University of California San Diego}

%
%

%

\begin{article}

%
%

\noindent\textbf{Contents of this file}
\begin{enumerate}
\item Text S1: Stochastic moist energy balance model derivation
\item Text S2: Surface effective heat capacity
\item Figure S1: Maps of temperature standard deviation in CESM
\item Figure S2: CESM zonal-mean temperature standard deviation over time
\item Figure S3: Maps of heat capacity
\item Figure S4: SMEBM with different warming patterns
\item Figure S5: Diagnosed noise from T-TOA regression
\item Figure S6: SMEBM with diagnosed noise from regression
\item Figure S7: Ensemble-mean temperature in CESM
\item Figure S8: Forced feedback parameter
\item Figure S9: Effective surface heat capacity from variability

\end{enumerate}



\begin{figure}[h]
\noindent\includegraphics[width=\textwidth]{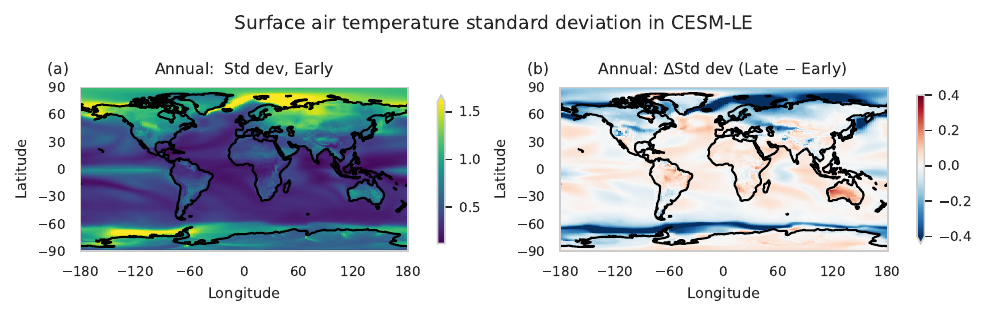}
\caption{Map of (unfiltered) temperature standard deviation on annual mean timescales in (a) the early 21st century (2000-2025), (b) the difference between the late (2075-2100) and early 21st century. Note that the zonal-mean of the temperature variability (which can be visually estimated here) is not equal to the variability of the zonal-mean temperature (which we focus on in this study).}
\label{fig:maps}
\end{figure}

\begin{figure}[h]
\noindent\includegraphics[width=\textwidth]{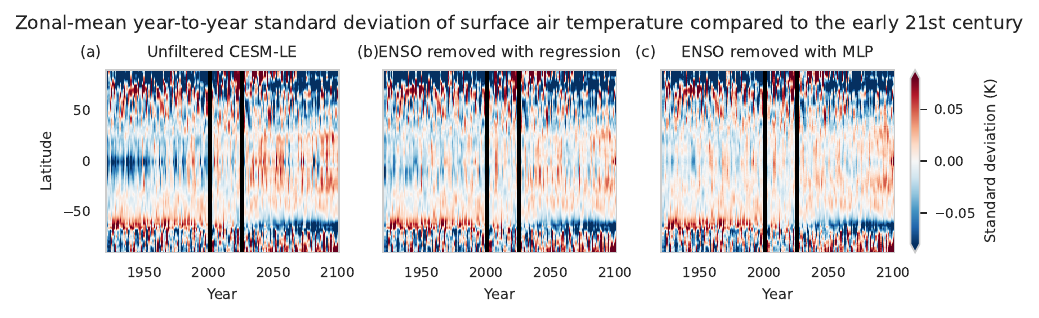}
\caption{Ensemble standard deviation of the zonal-mean surface air temperature shown relative to the mean standard deviation averaged over years 2000-2025 (averaging period denoted with the black lines). (a) using the original CESM-LE data with no processing, (b) using $A'_{ENSOfiltered}$, (c) using $A'_{ENSOfiltered,MLP}$.}
\label{fig:overtime}
\end{figure}

\begin{figure}[h!]
\noindent\includegraphics[width=\textwidth]{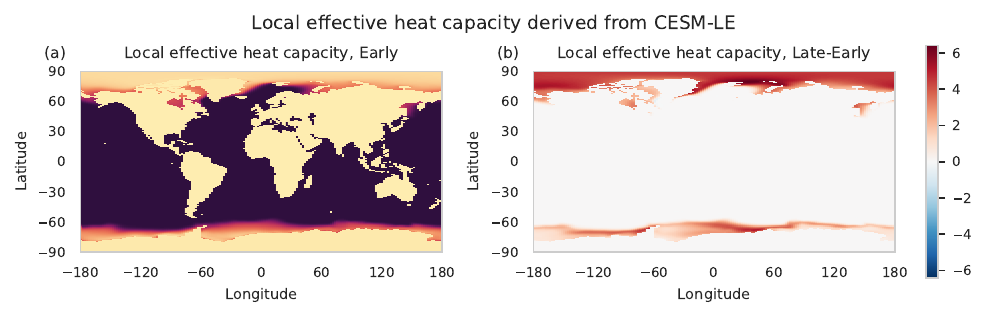}
\caption{Maps of effective heat capacity in (a) the early 21st century, and (b) the difference between the end of the 21st century and the beginning. Figure 2a in the main text is the zonal-mean of the field plotted here.}
\label{fig:overtime}
\end{figure}
\clearpage

\clearpage

\noindent\textbf{Text S1: Stochastic moist energy balance model derivation} 

We consider the zonal-mean  energy budget for anomalies from a reference climate (denoted with primes) in a moist atmospheric column at a given latitude:
\begin{equation}
\left<\left< \frac{\partial h_{atm}'}{\partial t} \right>_p\right>_\theta = F_{TOA}'-F_{sfc}'+F_{AHT}'.
\label{eq1}
\end{equation}
Here, the prime indicates anomalies from the reference climate, $h_{atm}$ is the MSE at a given location in the atmospheric column, $\left<\,\right>_p$ indicates column integration defined as
$\left< \frac{\partial h_{atm}'}{\partial t} \right>_p \equiv  \int^{p_{sfc}}_{p_{TOA}} \,\frac{\partial h_{atm}'}{\partial t}\,\frac{dp}{g} , $
and $\left<\,\right>_\theta$ indicates a zonal average across all longitudes.
$F_{TOA}$ is the zonal-mean net radiative flux downwards at the top of the atmosphere, $F_{sfc}$ is the zonal-mean heat flux into the surface, and $F_{AHT}$ is the vertically-integrated zonal-mean convergence of atmospheric heat transport.

\textbf{TOA flux:} The TOA radiative anomaly is approximated to be proportional to the zonal-mean surface temperature anomaly $T'$ with a stochastic residual term:
\begin{equation}
F_{TOA}' = \lambda\,T' + \sigma_{\xi}\,\xi,
\label{eq1b}
\end{equation}
with $\lambda$ equal to the radiative feedback and $\sigma_{\xi}\,\xi$ a noise term  that can be viewed as the influence of weather fluctuations on the TOA flux. Note that with this sign convention, a negative $\lambda$ is stabilizing.

\textbf{Ocean mixed layer:} The change in the anomalous zonal-mean surface temperature anomaly $T'$ is approximated to be proportional to the anomalous heat uptake into the surface (e.g., the ocean mixed layer for regions over the ocean) from the atmosphere ($F_{sfc}'$) and from the zonal-mean ocean heat flux convergence ($F_q'$) for locations over the ocean:
\begin{equation}
C\,\frac{\partial T'}{\partial t} = F'_{sfc}+F'_q,
\end{equation}
with $C$ denoting the surface effective heat capacity. Here, we assume that the surface air temperature and ocean mixed layer are in equilibrium with each other, which is approximately applicable on interannual timescales. This assumption is the same as in the slab ocean model mode in Proistosescu et al. 2018.

\textbf{Diffusion of MSE:} In the standard MEBM framework, $F_{AHT}$ is parameterized as Fickian diffusion of the near-surface MSE assuming constant relative humidity (e.g., Hwang \& Frierson, 2010). This parameterization allows column energy to be expressed as a function of the near-surface air temperature, $T$. Hence, we let
\[
F_{AHT} =  D \, \nabla^2\,h(T).
 \]
Here, the near-surface MSE is
\[
{h}(T) \approx c_p \,T+L_v\,q(T),
\]
neglecting changes in gravitational potential energy, with the approximation that the relative humidity  remains constant. The specific humidity is approximated as
\[ q(T) \approx \frac{\epsilon\, r\,e_0}{p} \, \exp\bigg[\frac{a\,T}{b+T}\bigg],
\]
with parameter values as given in Beer \& Eisenman (2022). In this formalism the surface temperature $T$ is expressed as the departure from 273.15K. Unlike previous MEBM studies that have investigated
 long-term mean anomalies from a reference climate in response to greenhouse forcing, here we investigate time-evolving climate fluctuations. This leads to issues related to the nonlinear rectification of the fluctuations onto the long-term mean climate due to the nonlinearity  of ${h}(T)$. The rectification is expected to cause only a small correction, and we omit it by linearizing the MSE about the zonal-mean reference climate $T_0$ (similar to Merlis \& Henry 2018):
\[
h(T) \approx {h}(T_0) + \left[\frac{d {h}(T)}{d T}\right]_{T_0}\,T',
\]
since the surface temperature anomaly is $T' \equiv T-T_0$. Taken together, this implies that the anomaly in atmospheric heat transport convergence compared with the reference climate is
\begin{equation}
F_{AHT}' = D \, \nabla^2\, \left(\left[\frac{d {h}(T)}{d T}\right]_{T_0}\, T' \right). 
\label{MSE}
\end{equation}

\textbf{Governing equation:} Inserting Equations \eqref{eq1b}-\eqref{MSE}  into Equation \eqref{eq1} (with the assumption that $F_q'=0$) leads to:
\[
C\,\frac{\partial T'}{\partial t} + \left<\left< \frac{\partial h_{atm}'}{\partial t} \right>_p\right>_\theta = \lambda\,T' + D \, \nabla^2\, \left(\left[\frac{d {h}(T)}{d T}\right]_{T_0}\, T' \right)  + \sigma_\xi\,\xi.
\]

Since changes in  MSE storage in the atmospheric column tend to be considerably smaller than changes in surface (e.g., the ocean mixed layer) heat storage on year-to-year timescales, we neglect the second term on the left-hand side, leaving
\begin{equation}
C\,\frac{\partial T'}{\partial t}  = \lambda\,T' + D \, \nabla^2\, \left(\left[\frac{d {h}(T)}{d T}\right]_{T_0}\, T' \right)  + \sigma_\xi\,\xi .
\label{eq2}
\end{equation}

We note that this model is used for all latitudes and the assumption of larger surface heat capacity than atmospheric heat capacity holds less well in regions over the land and sea ice than over the ocean, which is a caveat of this formulation.

\textbf{Comments on noise in the SMEBM:} For simplicity we have taken $\xi$ to be noise that is white in time and space, scaled by a constant amplitude $\sigma_\xi$. The magnitude of $\sigma_\xi$ is tuned to give a reasonable match to the temperature standard deviation in the early period (beginning of the 21st century) and then left unchanged for all perturbation experiments. 

We assess the impact of noise with a spatial structure, by fitting $\sigma_{\xi}$ as the standard deviation of the residuals of the regression between temperature anomalies and TOA anomalies (the same regression through which we find $\lambda$ used in the model) rather than assuming its magnitude to be constant in space and between climates. The standard deviation of these residuals is shown in Figure \ref{fig:noise_regression} for both the early and late 21st century. We find that the noise from the TOA-temperature relationship has spatial structure, but it does not change substantially between climates. We show in Figure \ref{fig:noise_regression_results} that the approximate spatial structure of the resultant temperature variability in the SMEBM is similar when we use this noise structure and when we use a uniform noise magnitude (albeit with smaller temperature variability values). 

This highlights that despite neglecting spatial structure in noise intensity a simple test adding this does not significantly impact our results. However, other assumptions about the noise were not tested in these sensitivity studies and could be impactful. For example, we assumed that the noise in the system is from the T-TOA relationship rather than from other sources. Unaccounted for noise that could be present in other terms (e.g., atmospheric transport) could change the results here regarding the appropriate spatial structure of noise intensity and how important noise change is between climates (cf. Schneider et al. 2015). Additionally, factors such as spatial autocorrelation of the noise could be important and should be explored more, although simple tests adding this did not significantly change our results (not shown).

\begin{figure*}[h!]
  \noindent\includegraphics[width=\textwidth,angle=0]{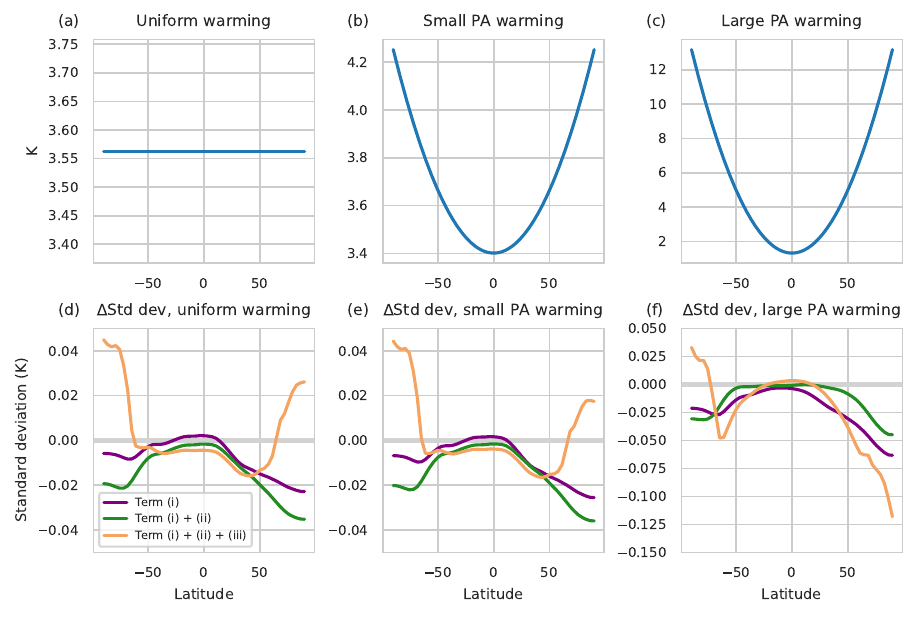}\\
  \caption{Variability in the SMEBM with different warming patterns, due just to the contribution of changed temperature (neglecting changed heat capacity due to sea ice retreat or local feedback changes, and thus analogous to Figure 3c in the main text). The first row shows the warming pattern, as (a) uniform warming, found from the global average of the CESM warming pattern (see Figure \ref{fig:CESMtemp}b), (b) an idealized warming pattern with the same global average but a small amount of polar amplification (polar warming is factor of 1.25 larger than equatorial warming), and (c) an idealized warming pattern with the same global average but a large amount of polar amplification (polar warming is factor of 10 larger than equatorial warming).  The second row shows the variability change after running the SMEBM with each warming pattern, decomposed into the three terms from Equation 7 in the main text. In addition to being sensitive to the warming pattern, the variability change is sensitive to the temperature prior to the temperature perturbation, which here is from the early 21st century in the CESM-LE in all panels (see Figure \ref{fig:CESMtemp}a).} \label{fig:sebm_designedwarming}
\end{figure*}


\begin{figure}[ht]
\centering
\noindent\includegraphics[width=\textwidth]{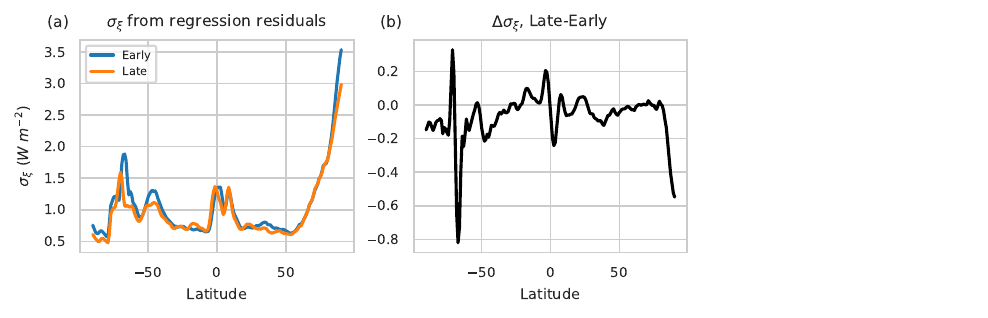}
\caption{$\sigma_{\xi}$ found from the standard deviation of the residuals in the regression between surface temperature and the TOA flux defined by Equation \ref{eq1b} for (a) each period, either early 21st century or late 21st century, and (b) the difference between the late and early periods.}
\label{fig:noise_regression}
\end{figure}

\begin{figure}[ht]
\centering
\noindent\includegraphics[width=\textwidth]{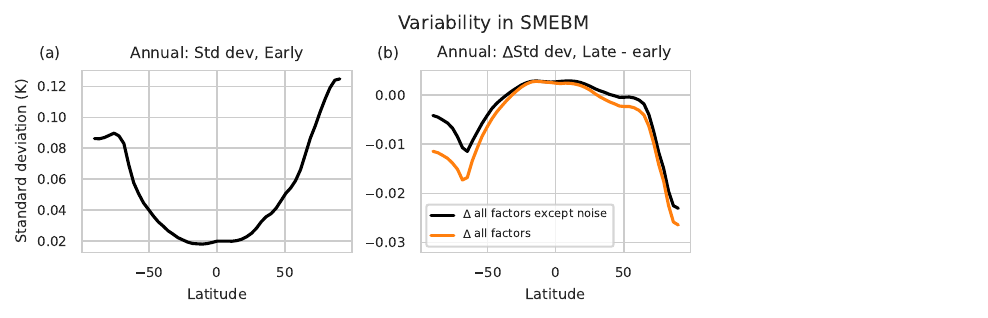}
\caption{Results when using $\sigma_{\xi}$ (found from the standard deviation of regression as in Figure \ref{fig:noise_regression}) as the noise intensity rather than assuming a spatially uniform magnitude in (a) the early period, and (b) the difference between the late and early periods. Panel (a) is comparable to Figure 3a in the main text, while the black line in panel (b) is comparable to the black line in Figure 3b. The orange line in panel (b) shows the additional impact of allowing the noise intensity to vary between climates (i.e., switching from the early to late noise profile in Figure \ref{fig:noise_regression}).}
\label{fig:noise_regression_results}
\end{figure}

\begin{figure}[ht]
\centering
\noindent\includegraphics[width=\textwidth]{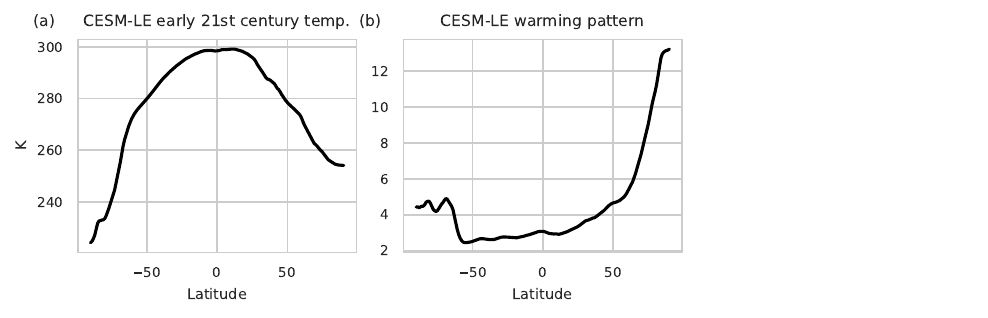}
\caption{The zonal-mean and ensemble-mean CESM-LE temperature (a) during the early 21st century (2000-2025) and (b) the difference between the late (2075-2100) and early 21st century.}
\label{fig:CESMtemp}
\end{figure}

\begin{figure}[ht]
\centering
\noindent\includegraphics[width=\textwidth]{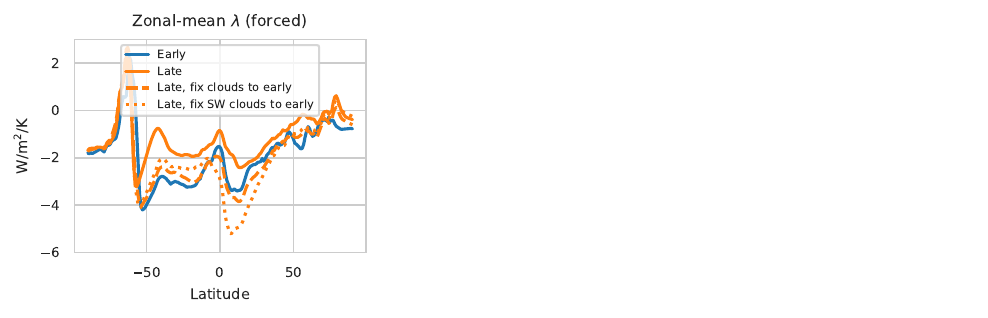}
\caption{The zonal-mean feedback parameter found through the ensemble-mean forced response and via the equation $ \lambda(x,t_1,t_2)=\sum_i \frac{\frac{\partial R_i}{\partial x_i}\Delta x_i}{\Delta T}$ where $\frac{\partial R}{\partial x}$ is the radiative kernel, $\Delta x$ is the change in a relevant climate variable, and $\Delta T$ is the local near-surface temperature change between times $t_1$ and $t_2$. The forced feedback is found for each period by comparing the time-mean and ensemble-mean variables to the 25 years immediately preceding the relevant period (e.g., for the early period, the reference climate $t_1$ is the average of years 1975-2000 and the perturbed climate $t_2$ is the average of years 2000-2025). The climate feedback is decomposed into components (Planck, lapse rate, albedo, shortwave/longwave water vapor, and shortwave/longwave cloud), with the total forced local radiative feedback being the sum over all components. In blue, we show the early period feedback and in orange we show the late period feedback. The dashed orange line shows fixing the cloud feedback to the early period while using the feedback computed from the late period for all other components and the dotted orange line shows the same but only fixing the shortwave cloud feedback. We used the CAM5 kernel and the cloud adjustment method contained in ClimKern, although we did not supply the radiative forcing as an input to the adjustment method due to lack of availability.}
\label{fig:feedback}
\end{figure}

\begin{figure}[ht]
\centering
\noindent\includegraphics[width=\textwidth]{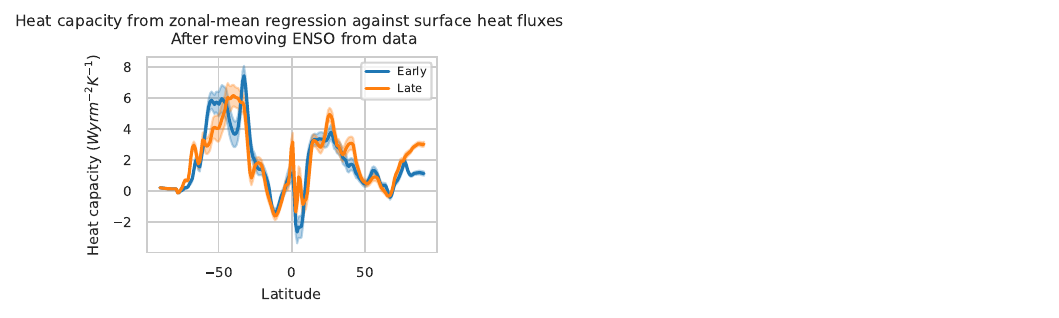}
\caption{The effective surface heat capacity fit from data using Equation (3) and the assumption that $F_q'=0$, as discussed in Text S2.}
\label{fig:feedback}
\end{figure}

\noindent\textbf{Text S2: Surface effective heat capacity} \\
We note that the effective heat capacity for the zonal-mean temperature is not necessarily the same as the zonal-mean of the effective heat capacity. These would only be equivalent in the limit of infinite coupling between longitudes, which can be readily verified using a two box model of the form $c_i dT_i/dt= \lambda T_i+D(T_j-T_i)+\eta_i$ for adjacent longitudes.

Our simplified assumption for the heat capacity in the zonal-mean SMEBM, where we used the zonal-mean of chosen heat capacities for the ocean, land, and sea ice should not be interpreted as implying that there is infinite coupling between longitudes. Rather than using fixed, physical heat capacities for each individual location, we view the effective heat capacity of each surface type as a tuneable parameter. Thus, our approach parametrizes the effective surface heat capacity of the zonal-mean temperature using the average of individual effective heat capacities depending on the surface type, assuming that the dominant variations in heat capacity between latitudes and climates are driven by different surface types.\\
\textbf{Fitting $C$ from data:}
We could have fit $C$ from flux data in the same way that we fit $\lambda$. For $C$, this process uses Equation (3), with the assumption that $F_q'=0$.

Similar to fitting $\lambda$, we take the anomalies from the ensemble mean of the zonal-mean near-surface temperature and the surface heat fluxes. We apply the linear ENSO filtering method to remove the impact of ENSO. We then take annual averages of the data and write $\frac{\partial T'}{\partial t}$ with centered differences, such that $\frac{\partial T'}{\partial t}_i=\frac{T_{i+1}-T_{i-1}}{2\Delta t}$, with $\Delta t=1$ year and the subscript $_i$ denoting the year.

We find $C$ in each 25 year period (either early or late 21st century) by regressing the annual average surface heat flux onto the estimated $\frac{\partial T'}{\partial t}_i$, yielding $C$ in each period in units of W yr m$^{-2}$ K$^{-1}$. The results are shown in Figure S9. 

We note that the negative heat capacity at the equator is likely due to the assumption that $F_q'=0$. Due to the missing variable $F_q'$ which biases the slopes, we do not use the fit $C$ from Figure S9 in the SMEBM. However, the fit $C$ from data adds evidence for our parameterized effective heat capacity used in this work (Figure 2a), as the largest changes in $C$ between the two time periods are at the poles.

\newpage

\end{article}
\clearpage


%
%
%
%